\documentstyle[prd,aps,preprint,epsfig,floats]{revtex}
\tightenlines

\def\dsodt{ds_{23}\over dt}
\def\dstdt{ds_{13}\over dt}
\def\dsthdt{ds_{12}\over dt}

\def\be{\begin{equation}}
\def\ee{\end{equation}}
\def\ba{\begin{eqnarray}}
\def\ea{\end{eqnarray}}
\def\br{\begin{array}}
\def\er{\end{array}}

\def\DESepsf(#1 width #2){\epsfxsize=#2 \epsfbox{#1}}

\begin{document}

\draft


\title{{\Large\bf Neutrino Mass and Grand Unification}}
\author{\bf  R.N. Mohapatra}

\address{ Department of Physics, University of Maryland, College Park,
MD-20742, USA}
\date{December, 2004}
\maketitle

\begin{abstract}

Seesaw mechanism appears to be the simplest and most appealing
way to understand small neutrino masses observed in recent
experiments. It introduces three right handed neutrinos with
heavy masses to the standard model, with at least one mass required
by data to be close to the scale of conventional grand unified
theories.  This may be a hint that the new
physics scale implied by neutrino masses and grand unification of
forces are one and the same. Taking this point of view seriously,
I explore different ways to resolve the puzzle of large neutrino
mixings in grand unified theories such as SO(10) and models based on its
subgroup $SU(2)_L\times SU(2)_R\times SU(4)_c$.

\end{abstract}

\vskip1.0in
\newpage

\section{Introduction}
The discovery of neutrino masses and mixings has been an important
milestone in the history of particle physics and rightly
qualifies as the first evidence for new physics beyond the
standard model. The amount of new information on neutrinos
already established from various oscillation experiments has
provided very strong clues to new symmetries of particles and
forces and new directions for unification. Enough puzzles have
emerged making this field a hotbed for research with implications
ranging from ideas such as supersymmetry and grand unification to
cosmology and astrophysics.

A major cornerstone for the theory research in this field has
been the seesaw mechanism introduced in the late seventies
\cite{seesaw} to understand why neutrino masses are so much
smaller than the masses of other fermions of the standard model.
Even though there was no evidence for neutrino masses then, there
were very well motivated extensions of the standard models that
led to nonzero masses for neutrinos. It was therefore incumbent
on those models that they have a mechanism for understanding why
upper limits on neutrino masses known at that time were so small.
Seesaw mechanism introduces three right handed neutrinos into the
standard model
with very large Majorana masses and predicts that observed neutrinos 
are their own anti-particles. A very appealing aspect of this mechanism
is not only the beauty and elegance it brings to the standard model by
restoring quark-lepton symmetry but also the new insight it
provides into such questions as the origin of parity violation
and Dirac vrs Majorana nature of the neutrino.

The first conclusive evidence for nonzero neutrino masses appeared
in 1998. During the past six years, we have learnt that neutrinos
not only have mass but they also mix among themselves with a
pattern which is very different from that among quarks. The
equation below summarizes our present knowledge about neutrino
masses and mixings\cite{rev}  in the notation $|\nu_\alpha>= \sum
U_{\alpha i}|\nu_i>$ (where $\alpha= e,\mu.\tau$ is the flavor index and
$i=1,2,3$ denotes the mass eigenstate index). For the CP conserving case
$U_{\alpha i}$ are functions
of three angles, $\theta_{ij}$ and for these angles we
have:
\begin{eqnarray}
ain^22\theta_A\equiv sin^22\theta_{23}\geq 0.89\\ \nonumber
\Delta m^2_A\simeq 1.4\times 10^{-3}~ eV^2-3.3\times 10^{-3}~ eV^2\\
\nonumber
sin^2\theta_\odot\equiv sin^2\theta_{12}\simeq 0.23-0.37\\ \nonumber
\Delta m^2_\odot\simeq 7.3\times 10^{-5}~ eV^2-9.1\times 10^{-5}~ eV^2\\
\nonumber
sin^2\theta_{13}\leq 0.047
\end{eqnarray}
For the sake of comparision, note the corresponding quark
mixing angles i.e. $\theta^q_{12}\simeq 0.22$;
$\theta^q_{23}\simeq 0.04$ and $\theta^q_{13}\simeq 0.004$. Clearly,
the mixing pattern in the lepton sector is very different from that
among quarks.

It is also important to point that while the mass differences
among neutrinos are fairly well determined, the situation with
respect to absolute values of masses is far from clear. This is
another major gap in our understanding of neutrinos compared to
quarks. At present, there are three equally viable mass
arrangements among the neutrinos:
\begin{itemize}

\item (i) Normal hierarchy i.e. $m_1\ll m_2 \ll m_3$. In this case,
we can deduce the value of $m_3 \simeq \sqrt{\Delta m^2_{23}}
\equiv \sqrt{\Delta m^2_A}\simeq 0.03-0.07$ eV. In this case $\Delta
m^2_{23}\equiv m^2_3-m^2_2 > 0$.
 The solar neutrino oscillation involves the two lighter levels. The mass
of the lightest neutrino is unconstrained. If $m_1\ll m_2$, then we get
the value of $m_2 \simeq \simeq 0.008$ eV.

\item (ii) Inverted hierarchy i.e. $m_1 \simeq m_2 \gg m_3$ with
$m_{1,2} \simeq \sqrt{\Delta m^2_{23}}\simeq 0.03-0.07$ eV. In this case,
solar neutrino oscillation takes place between the heavier levels and we
have $\Delta m^2_{23}\equiv m^2_3-m^2_2 < 0$.

\item (iii) Degenerate neutrinos i.e. $m_1\simeq m_2 \simeq m_3$.

\end{itemize}
 There are a large number of experiments in the planning
stage to improve our knowledge of mixings, to determine the mass ordering 
and also to find out whether neutrinos are Majorana (i.e. their
own antiparticles) or Dirac fermions. These are not only crucial
pieces of information about the neutrinos that we need to know to
elevate our knowledge of them to the same level as the quarks but
it is becoming increasingly clear that they will also point very
clearly to the direction of new physics beyond the standard model.
For instance if neutrinos are established to be Dirac fermions, seesaw
mechanism in its simplest form will not be able to describe their masses 
and a major theoretical idea will be disproved.

If we accept the seesaw mechanism as the explanation for the smallness of
neutrino masses, the next major challenge for theory is to understand
the unusual mixing pattern among them. The hope is that in the process of
understanding the mixings we will find out which of the mass patterns is
realized in Nature and more importantly, will get a definite clue to
the nature of new physics.

 In this talk I will give some promising possibilities for this new
physics and discuss their experimental tests. In particular, I
will argue that the seesaw mechanism for small neutrino masses
requires a scale of new physics close to the traditional scale of
grand unification where all forces and matter are supposed to
become unified and a new symmetry B-L which naturally arises if
the gauge group is assumed to be SO(10)\cite{so10}. I will then
show that a minimal version of supersymmetric SO(10) provides a
very natural way to understand the large solar as well as
atmospheric neutrino mixing angles while predicting a value for
the mixing angle $\theta_{13}\sim 0.1- 0.18$ depending on details.
This prediction can be tested by the various planned
reactor\cite{reactor} and long baseline experiments\cite{lbl}. I
will also discuss two other related ideas which are outside the
SO(10) framework but are based on one of the maximal subgroups of
SO(10) i.e. $SU(2)_L\times SU(2)_R \times SU(4)_c$\cite{ps} that
also unifies quarks and leptons and then argue that measurement
of the parameter $\theta_{13}$ may provide crucial insight into
the question of whether there is quark-lepton unification at high
scale.

While in this talk I will assume that there are only three
neurtinos, we do not know for sure how many neutrinos there are. In
particular if the LSND results are confirmed by the Mini Boone
experiment\cite{janet}, we will have evidence that there are more
neutrinos and the discussions presented here will have to be extended.

 \section{Seesaw mechanism, B-L and left-right symmetry}
In order to introduce the seesaw mechanism, which will form the
anchor for the main body of the talk, let us start with a
discussion of neutrino mass in the standard model. It is based on
the gauge group $SU(3)_c\times SU(2)_L\times U(1)_Y$ group under
which the quarks and leptons transform as follows:
 Quarks: $Q_L(3,2, {1\over 3})$; $u_R(3, 1, {4\over 3})$;
   $ d_R(3, 1,-\frac{2}{3})$;  Leptons $L(1, 2 -1)$;  $e_R(1,1,-2)$;
Higgs Boson ${\bf H}(1, 2, +1)$; Gluons  $G_a(8, 1, 0)$ and Weak
Gauge Fields  $W^\pm, Z, \gamma$. The electroweak symmetry
$SU(2)_L\times U(1)_Y$ is broken by the vacuum expectation of the
Higgs doublet $<H^0>=v_{wk}\simeq 180$ GeV, which gives mass to
the gauge bosons and all fermions except the neutrino. The model
had been a complete success in describing all known low energy
phenomena, until the evidence for neutrino masses appeared.

Note that there is no right handed neutrino in the standard model
and this directly leads to massless neutrinos at the tree level.
The situation remains the same not only to all orders in
perturbation theory but also when nonperturbative effects are
taken into account. This is due to existence of an exact B-L
symmetry in the theory and the absence of the right handed
neutrino, $N_R$. The absence of the right handed neutrino from
the standard model of course destroys the symmetry between quarks
and leptons that is so obvious in weak interactions.

Once the right handed neutrinos ($N_R$) are included in the
standard model, new Yukawa couplings of the form
$h_\nu\bar{L}HN_R$ are allowed which after electroweak symmetry
breaking lead to a neutrino mass, $M_D\equiv h_\nu v_{wk}$.
Since $h_\nu$ is expected to be of same order as the charged
fermion Yukawa couplings in the model, these masses are much too
large to describe neutrino oscillations. Luckily, since the
$N_R$'s are singlets under the standard model gauge group, they
are allowed to have Majorana masses unlike the charged fermions.
We denote them by $M_RN^T_RC^{-1}N_R$ (where $C$ is the Dirac
charge conjugation matrix). The masses $M_R$ are not constrained
by the gauge symmetry and can therefore be arbitrarily large
(i.e. $M_R \gg h_\nu v_{wk}$). This together with mass induced by
Yukawa couplings (called the Dirac mass) leads to a the mass
matrix for the neutrinos (left and right handed neutrinos
together) which has the form
\begin{eqnarray}
{\cal M}_\nu~=~\pmatrix{0 & M_D\cr M^T_D & M_R}
\end{eqnarray}
where $M_D$ and $M_R$ are $3\times 3$ matrices. Diagonalizing
this mass matrix, one gets the mass matrix for the light
neutrinos (the seesaw formula) as:
\begin{eqnarray}
{\cal M}_\nu~=~-M^T_DM^{-1}_RM_D
\end{eqnarray}
Since as already noted $M_R$ can be much larger than $M_D$, one
finds that $m_\nu \ll m_{e,u,d}$ very naturally.

Seesaw mechanism of course raises its own questions:

\begin{itemize}

\item  Is there a natural reason for the existence of the right handed
neutrinos other than quark-lepton symmetry ?

\item What determines the scale of $M_R$ ?

\item  Is the seesaw mechanism by itself enough to explain all aspects of
neutrino masses and mixings ?

\end{itemize}

Below, we try to answer some of these questions. Restoration of
quark-lepton symmetry and unification of quarks and leptons
within a single gauge theory framework provided the first inspiration to
bring the right handed neutrino into particle physics\cite{ps}.
It is easy to see that in the presence of the $N_R$'s, the minimal
anomaly free gauge group of weak interactions expands beyond the
standard model and becomes the left-right symmetric group
$SU(2)_L\times SU(2)_R\times U(1)_{B-L}$\cite{lrs} which is a
subgroup of the $SU(2)_L\times SU(2)_R\times SU(4)_c$ group. This
makes the weak interactions parity conserving at short
distances\cite{lrs}, providing another appealing feature of adding the
right handed neutrino. 
To see this explicitly, we give in Table I, the assignment
of fermions and Higgs fields to the left-right gauge group.
\begin{center}
{\bf Table I}
\end{center}

\begin{center}
\begin{tabular}{|c|c|} \hline\hline
Fields           & SU$(2)_L \, \times$ SU$(2)_R \, \times$ U$(1)_{B-L}$ \\
                 & representation \\ \hline
$Q_L\equiv \pmatrix{u_L \cr d_L}$     & (2,1,$+ {1 \over 3}$) \\
$Q_R \equiv \pmatrix{u_R \cr d_R}$            & (1,2,$ {1 \over 3}$) \\
$L_L\equiv \pmatrix{\nu_L \cr e_L}$                & (2,1,$- 1$) \\
$L_R\equiv \pmatrix{\nu_R \cr e_R}$            & (1,2,$- 1$) \\
$\phi$     & (2,2,0) \\
$\Delta_L$         & (3,1,+ 2) \\
$\Delta_R$       & (1,3,+ 2) \\
\hline\hline
\end{tabular}
\end{center}

It is clear that this theory leads to a weak interaction Lagrangian of the
form
\begin{eqnarray}
{\cal L}_{wk}~=~ \frac{g}{2}\left(\vec{j}^\mu_L\cdot \vec{W}_{L,\mu}
+\vec{j}^\mu_R\cdot \vec{W}_{R,\mu}\right)
\end{eqnarray}
which is parity conserving prior to symmetry breaking.
 Furthermore, the electric charge formula is given by\cite{marshak}:
\begin{eqnarray}
Q~=~I_{3L}~+~I_{3R}~+~\frac{B-L}{2}.
\end{eqnarray}
where all the terms have physical meaning unlike the
case of the standard model.

The left-right symmetric theories face two challenges: (i) how
does the predominantly V-A nature of weak interactions emerge in
such a theory and (ii) how does one understand the small neutrino
masses since $SU(2)_R$ makes both the electron and the neutrino
much more similar than they were in the standard model. We will
see that both these challenges are met in one stroke i.e.
breakdown of $SU(2)_R\times U(1)_{B-L}$ symmetry to $U(1)_Y$ not
only explains the V-A nature of weak interactions but it also
explain why $m_\nu \ll m_e$ via the seesaw mechanism. The seesaw
scale then becomes the scale of parity violation.
 Furthermore, when the
gauge symmetry $SU(2)_R\times U(1)_{B-L}$ is broken down while keeping the
standard model symmetry unbroken, one finds from Eq. (5) the relation
$\Delta
I_{3R}~=-\Delta\frac{B-L}{2}$. This connects $B-L$ breaking
 to the breakdown of parity symmetry i.e. $\Delta I_{3R}\neq 0$ and
clearly implies that neutrinos must be Majorana particles.

To see this explicitly, we break the gauge symmetry of the
left-right model in two stages : in stage I, vacuum expectation
values (vev) of the Higgs multiplets $ \Delta_R(1,3,2)$ breaks the
left-right gauge symmetry to the standard model gauge group and in
stage II by the bidoublet $\phi(2,2,0)$ vev breaks the standard
model group to $SU(3)_c\times U(1)_{em}$. In the first stage of
symmetry breaking, the right handed neutrino picks up a mass of
order $f<\Delta^0_R>\equiv fv_R$. Denoting the left and right
handed neutrino by $(\nu, N)$ (in a two component notation), the
mass matrix for neutrinos at this stage looks like
\begin{eqnarray}
{\cal M}^0_\nu~=~\pmatrix {0 & 0 \cr 0 & fv_R}
\end{eqnarray}
At this stage, familiar standard model particles remain massless.
As soon as the standard model symmetry is broken by the bidoublet
$\phi$ i.e. $<\phi>\equiv \pmatrix{\kappa & 0\cr 0 & \kappa'}$,
the W and Z boson as well as the fermions pick up mass. I will
generically denote $\kappa,\kappa'$ by a common symbol $v_{wk}$.
The contribution to neutrino mass at this stage look like
\begin{eqnarray}
{\cal M}^0_\nu~=~\pmatrix {fv_L & h v_{wk} \cr hv_{wk} & fv_R}
\end{eqnarray}
Note the appearance of a new term in the neutrino mass matrix
 i.e. $v_L~=~\frac{v^2_{wk}}{v_R}$ compared to the seesaw matrix given in
 Eq. (1). This is a reflection of parity
invariance of the model. Diagonalizing this matrix, we get a
modified seesaw formula for the light neutrino mass matrix
\begin{eqnarray}
{\cal M}_\nu~=~fv_L-h^T_\nu f^{-1}_Rh_\nu
\left(\frac{v^2_{wk}}{v_R}\right)
\end{eqnarray}
 The important point to note is that $v_L$ is
suppressed by the same factor as the second term so that despite
the new contribution to neutrino masses, seesaw suppression
remains\cite{seesaw2}. This is called the type II seesaw in
contrast with the formula in Eq. (2) which is called type I seesaw
formula.

An important physical meaning of the seesaw formula is brought out when
it is viewed in the context of left-right models. Note that
$m_\nu\rightarrow 0$ when $v_R$ goes to infinity. In the same
limit the weak interactions become pure V-A type. Therefore, left-right
model derivation of the seesaw formula smoothly connects smallness of
neutrino mass with suppression of V+A part of the weak interactions
providing an important clarification of a major puzzle of the standard
model i.e. why are weak interactions are near maximally parity violating ?
The answer is that they are near maximally parity violating because the
neutrino mass happens to be small.

In a subsequent section, we will discuss the connection of the
seesaw mass scale with the scale of grand unification. SO(10) is
the simplest gauge group that contains the right handed neutrino
needed to implement the seesaw mechanism and also it is important
to note that the left-right symmetric gauge group is a subgroup
of the SO(10) group, which therefore provides an attractive over
all grand unified framework for the discussion of neutrino
masses. The extra bonus one may expect is that since bigger
symmetries tend to relate different parameters of a theory, one
may be able to predict neutrino masses and mixings. We will
present a model where indeed this happens.

Before proceeding further, it is important to point out that type I and
type II seesaw can be tested by the nature of neutrino spectrum in a model
independent way. Since type I seesaw involves the Dirac mass of the
neutrino, a general expectation is that it scales with generation the same
way as the charged fermions of the standard model. In this case, unless
there is extreme hierarchy among the right handed neutrinos, one would
expect the spectrum to be hierarchical. On the other hand, it has been
realized for a long time\cite{caldwell} that if neutrino masses are
quasi-degenerate, it is a tell-tale sign of type II seesaw with the
triplet vev term being the dominant one. However, a normal hierarchy can
also arise with type II seesaw as we discuss in the example
below. Therefore, whereas a normal hierarchy cannot distinguish between
type I and type II seesaw, a quasi-degenerate spectrum is a definite sign
of type II kind.

 \section{Seesaw and large neutrino mixings}
While seesaw mechanism provides a simple framework for
understanding the smallness of neutrino masses, it does not throw
any light on the question of why neutrino mixings are large. The
point is that mixings are a consequence of the structure of the
light neutrino mass matrix and the seesaw mechanism is only
statement about the scale of new physics. This can also be
understood by doing a simple parameter counting. If we work in a
basis where the right handed neutrino masses are diagonal, there
are 18 parameters describing the seesaw formula for neutrino
masses - three RH neutrino masses and 15 parameters in the Dirac
mass matrix. On the other hand, there are only nine observables
(three masses, three mixing angle and three phases) describing low
energy neutrino sector. Thus there are twice as many parameters
as observables. As a result, understanding neutrino mixings needs
inputs beyond the simple seesaw mechanism to fix the neutrino mass
matrix. Nonetheless, since the large mixings could arise from the
physics involving the seesaw formula e.g. flavor structure of
$M_R$, the large mixings are not in obvious contradiction with
quark lepton unification. This becomes clear in the examples
given below.

 Many seesaw models for large mixings have been considered in the
 literature\cite{king}. In the following section, I
will focus on a recently discussed minimal SO(10) model, where
without any assumption other than SO(10) grand unification, one
can indeed predict all but one neutrino parameters. I will then
consider a case where assumption of quasi-degeneracy in the
neutrino spectrum at high scale leads in a natural way via
radiative corrections to large mixings at low energies as well as
briefly describe a model of quark-lepton complementarity.

To understand the fundamental physics behind neutrino mixings, we
first write down the neutrino mass matrix that leads to maximal
solar and atmospheric mixing for the case of normal hierarchy:
\begin{eqnarray}
{\cal M}_\nu~=~\frac{\sqrt{\Delta m^2_A}}{2}\pmatrix{c\epsilon
&b\epsilon &d\epsilon\cr b\epsilon & 1+a\epsilon & -1 \cr
d\epsilon & -1 & 1+\epsilon}
\end{eqnarray}
where $\epsilon \simeq \sqrt{\frac{\Delta m^2_\odot}{\Delta
m^2_A}}$ and parameters $a,b,c,d$ are of order one. Any theory of
neutrino which attempts to explain the observed mixing pattern
for the case of normal hierarchy must strive to get a mass matrix
of this form.

It is important to point out that  the above mass matrix when
$a=1$ and $b=d$, becomes symmetric under the interchange of $\mu$
and $\tau$ and yields $\theta_{13}=0$. It was shown in two recent
papers that\cite{mutau}, if $\mu-\tau$ symmetry is broken via
$a\neq 1$ with $b=d$, then typically $\theta_{13}\sim
{\frac{\Delta m^2_\odot}{\Delta m^2_A}}$ whereas if we have
$b\neq d$, one gets $\theta_{13}\sim \sqrt{\frac{\Delta
m^2_\odot}{\Delta m^2_A}}$. It turns out that most grand unified
(or quark-lepton unified ) theories lead to $\theta_{13}\sim
\sqrt{\frac{\Delta m^2_\odot}{\Delta m^2_A}}$ (see examples
below). Therefore, measurement of the mixing parameter
$\theta_{13}$ may provide a way to test for possible quark-lepton
unification at high scales.

\section{A predictive minimal SO(10) theory for neutrinos}

The main reason for considering SO(10) for neutrino masses is that its
{\bf 16} dimensional spinor representation consists of all fifteen
standard model fermions plus the right handed neutrino arranged according
to the it $SU(2)_L\times SU(2)_R\times SU(4)_c$\cite{ps} subgroup as
follows:
\begin{eqnarray}
{\bf \Psi}~=~\pmatrix{u_1 & u_2 & u_3 & \nu\cr d_1 & d_2 & d_3& e}
\end{eqnarray}
There are three such spinors for three fermion families.

In order to implement the seesaw mechanism in the SO(10) model,
one must break the B-L symmetry. In supersymmetric SO(10) models,
how B-L breaks has profound consequences for low energy physics.
For instance, if B-L is broken by a Higgs field belonging to the
{\bf 16} dimensional Higgs field (to be denoted by $\Psi_H$),
then the field that acquires a nonzero vev has the quantum
numbers of the $\nu_R$ field i.e. B-L breaks by one unit. In this
case higher dimensional operators of the form
$\Psi\Psi\Psi\Psi_H$ will lead to R-parity violating operators in
the effective low energy MSSM theory such as $QLd^c, u^cd^cd^c$
etc which can lead to large breaking of lepton and baryon number
symmetry and hence unacceptable rates for proton decay. This
theory also has no dark matter candidate.

On the other hand, if one breaks B-L by a {\bf 126} dimensional
Higgs field, the member of this multiplet that acquires vev has
$B-L=2$.  R-parity is therefore left as an automatic symmetry of
the low energy Lagrangian.
 There is a naturally stable dark matter in this case.
It has recently been shown that this class of models lead to a very
predictive scenario for neutrino mixings\cite{babu,last,goran,goh}. We
summarize this model below.

As already noted earlier, any theory with asymptotic parity
symmetry leads to type II seesaw formula and if B-L is broken by a
{\bf 126} field, then the first term in the type II seesaw
formula can in principle dominate in the seesaw formula. We will
discuss a model of this type below.

The basic ingredients of this model are that one considers only two Higgs
multiplets that contribute to fermion masses i.e. one
{\bf 10} and
one {\bf 126}. A unique property of the {\bf 126}
multiplet is that it not only breaks the B-L symmetry and therefore
contributes to
right handed neutrino masses, but it also contributes to charged fermion
masses by virtue of the fact that it contains MSSM doublets which mix with
those from the {\bf 10} dimensional multiplets and survive down to the
MSSM scale. This leads to a tremendous reduction of  the number of
arbitrary parameters, as we will see below.

There are only two Yukawa coupling matrices in this model: (i)
$h$ for the {\bf 10} Higgs and (ii) $f$ for the {\bf 126} Higgs.
SO(10) has the property that the Yukawa couplings involving the
{\bf 10} and {\bf 126} Higgs representations are symmetric.
Therefore if we assume that CP violation arises from other
sectors of the theory (e.g. squark masses) and work in a basis
where one of these two sets of Yukawa coupling matrices is
diagonal, then it will have only nine coupling parameters. Noting
the fact that the (2,2,15) submultiplet of {\bf 126} has a pair of
standard model doublets that contributes to charged fermion
masses, one can write the quark and lepton mass matrices as
follows\cite{babu}:
\begin{eqnarray}
M_u~=~ h \kappa_u + f v_u \\\nonumber
M_d~=~ h \kappa_d + f v_d \\  \nonumber
M_\ell~=~ h \kappa_d -3 f v_d \\  \nonumber
M_{\nu_D}~=~ h \kappa_u -3 f v_u \\\nonumber
\end{eqnarray}
where $\kappa_{u,d}$ are the vev's of the up and down standard
model type Higgs fields in the {\bf 10} multiplet and $v_{u,d}$
are the corresponding vevs for the same doublets in {\bf 126}.
The vevs added to the Yukawa couplings give a total of 13
parameters in the theory. They are determined by 13 inputs (six
quark masses, three lepton masses and three quark mixing angles
and weak scale). There is therefore no free parameter in the
neutrino sector except for an overall seesaw scale.

 To determine the light neutrino masses, we use the seesaw
formula in Eq. (7), where the {\bf f} is nothing but the {\bf 126}
Yukawa coupling.  These models were extensively discussed in the
last decade\cite{last} using type I seesaw formula. It was
pointed out in Ref.\cite{goran} that if the direct triplet term
in type II seesaw dominates, then it provides a very natural
understanding of the large atmospheric mixing angle for the case
of two generations without invoking any symmetries. Subsequently
it was shown in Ref.\cite{goh} that the same $b-\tau$ mass
convergence also provides an explanation of large solar mixing as
well as small $\theta_{13}$ making the model realistic and
experimentally interesting.

 A simple way to see how large mixings arise in this model is to note that
when the triplet term dominates the seesaw formula,  we have the
neutrino mass matrix ${ M}_\nu \propto f$, where $f$ matrix is
the {\bf 126} coupling to fermions discussed earlier. Using the
above equations, one can derive the following sumrule :
\begin{eqnarray}
{ M}_\nu~=~ c (M_d - M_\ell)
\label{key}
\end{eqnarray}
To see how this leads to large atmospheric and solar mixing,
let us work in the basis where the down quark mass matrix is diagonal. All
the quark mixing effects are then in the up quark mass matrix i.e.
$M_u~=~U^T_{CKM}M^d_u U_{CKM}$. Note further that the minimality of the
Higgs content leads to the following sumrule among the mass matrices:
\begin{eqnarray}
k \tilde{M}_{\ell}~=~r\tilde{ M}_d +\tilde{ M}_u
\end{eqnarray}
where the tilde denotes the fact that we have made the mass
matrices dimensionless by dividing them by the heaviest mass of
the species i.e. up quark mass matrix by $m_t$, down quark mass
matrix by $m_b$ etc. $k,r$ are functions of the symmetry breaking
parameters of the model. Using the hierarchical pattern of quark
mixings, we can conclude that that we have
\begin{eqnarray}
M_{d,\ell}~\approx ~m_{b,\tau}\pmatrix{\lambda^3 & \lambda^3
&\lambda^3\cr \lambda^3 & \lambda^2& \lambda^2 \cr \lambda^3 & \lambda^2 &
1}
\end{eqnarray}
where $\lambda \sim 0.22$ and the matrix elements are supposed to
give only the approximate order of magnitude. An important
consequence of the relation between the charged lepton and the
quark mass matrices in Eq. (12) is that the charged lepton
contribution to the neutrino mixing matrix i.e. $U_\ell \simeq
{\bf 1} + O(\lambda)$ or close to identity matrix. As a result
the neutrino mixing matrix is given by
$U_{PMNS}~=~U^{\dagger}_\ell U_\nu \simeq U_\nu$. Thus the
dominant contribution to large mixings will come from $U_\nu$,
which in turn will be dictated by the sum rule in Eq. (11).

To show that $U_\nu$ has two large mixings, we extrapolate the
quark masses to the GUT scale and use the well known fact  that
$m_b-m_\tau \approx m_{\tau}\lambda^2$ for a wide range of values
of tan$\beta$. Using this the neutrino mass matrix
$M_\nu~=c(M_d-M_\ell)$ roughly takes the form
\begin{eqnarray}
M_{\nu}~=c(M_d-M_\ell)\approx ~m_0\pmatrix{\lambda^3 & \lambda^3
&\lambda^3\cr \lambda^3 & \lambda^2 & \lambda^2 \cr \lambda^3 & \lambda^2
& \lambda^2}
\end{eqnarray}
This mass matrix is in the form discussed in Eq. (8) and it is
easy to see that both the $\theta_{12}$ (solar angle) and
$\theta_{23}$ (the atmospheric angle) are now large. The detailed
magnitudes of these angles of course depend on the details of the
quark masses at the GUT scale. Using the extrapolated values of
the quark masses and mixing angles to the GUT scale, the
predictions of this model for various oscillation parameters are
given in Fig. 1,2 and 3 in a self expalanatory notation. The
predictions for the solar and atmospheric mixing angles fall
within 3 $\sigma$ range of the present central values. Note
specifically the prediction in Fig. 3 for $U_{e3}\simeq 0.18$
which can be tested in MINOS as well as other planned Long Base
Line neutrino experiments such as Numi-Off-Axis, JPARC etc. This model has
been the subject of many investigations, which we do not discuss
here\cite{more}.

\begin{figure}
\begin{center}
\epsfxsize8cm\epsffile{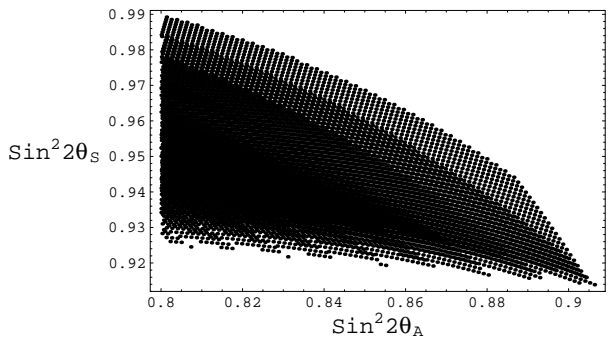}
\caption{
The figure shows the predictions of the minimal SO(10) model for
$sin^22\theta_{\odot}$ and
$sin^22\theta_A$ for the presently range of quark masses. Note that
$sin^22\theta_{\odot}\geq 0.9$ and $sin^22\theta_A\leq 0.9$
\label{fig:cstr1}}
\end{center}
\end{figure}

\begin{figure}
\begin{center}
\epsfxsize8cm\epsffile{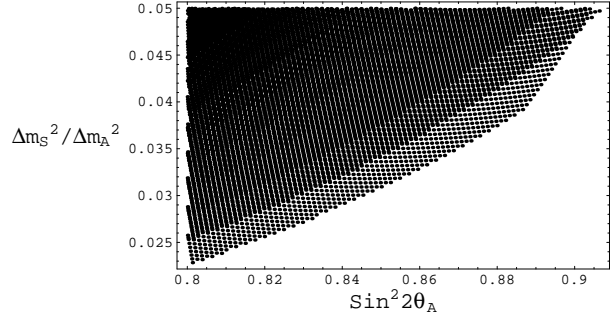}
\caption{
The figure shows the predictions of the minimal SO(10) model for
$sin^22\theta_{A}$ and
 $\Delta m^2_{\odot}/\Delta m^2_{A}$ for the range of quark masses and
mixings that fit charged lepton masses.
\label{fig:cstr2}}
\end{center}
\end{figure}

\begin{figure}
\begin{center}
\epsfxsize8cm\epsffile{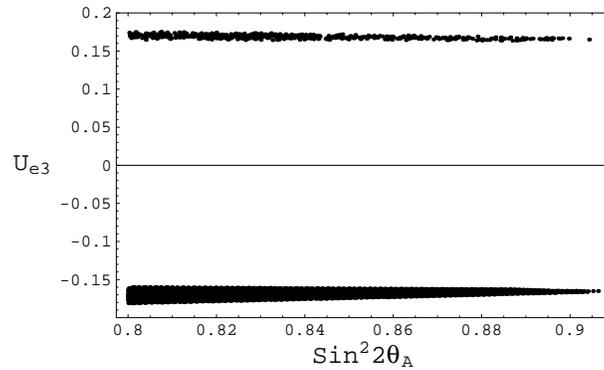}
\caption{
The figure shows the predictions of the minimal SO(10) model for
$sin^22\theta_{A}$
and $U_{e3}$ for the allowed range of parameters in the model. Note that
$U_{e3}$ is very close to the upper limit allowed by the existing reactor
experiments.
\label{fig:cstr3}}
\end{center}
\end{figure}

\subsection{CP violation in the minimal SO(10) model} In the
discussion given above, it was assumed that CP violation is
non-CKM type and resides in the soft SUSY breaking terms of the
Lagrangian. The overwhelming evidence from experiments seem to be
that CP violation is perhaps is of CKM type. It has recently been
pointed out that with slight modification, one can include CKM CP
violation in the model\cite{mimura}. The basic idea is to include
all higher dimensional operators of type $h'{\Psi \Psi}
\bar{\Delta}{\Sigma}/M$ where $\bar{\Delta}$ and $\Sigma$ denote
respectively the {\bf 126} and the {\bf 210} dimensional
representation. It is then clear that those operators
transforming as {\bf 10} and {\bf 126} representations will
simply redefine the $h,f$ coupling matrices and add no new
physics. On the other hand the higher dimensional operator that
transforms like an effective {\bf 120} representation will add a
new piece to all fermion masses. Now suppose we introduce a
parity symmetry into the theory which transforms $\Psi$ to
${\Psi^c}^*$, then it turns out that the couplings $h$ and $f$
become real and symmetric matrices whereas the {\bf 120} coupling
(denoted by $h'$) becomes imaginary and antisymmetric. This
process introduces three new parameters into the theory and the
charged fermion masses are related to the fundamental couplings
in the theory as follows:
\begin{eqnarray}
M_u~=~ h \kappa_u + f v_u~+ h'v_u \\  \nonumber
M_d~=~ h \kappa_d + f v_d ~+h'v_d\\  \nonumber
M_\ell~=~ h \kappa_d -3 f v_d -3h'v_d\\  \nonumber
M_{\nu_D}~=~ h \kappa_u -3 f v_u -3h'v_u\\
\end{eqnarray}
Note that the extra contribution compared to Eq. (10) is antisymmetric
which
therefore does not interfere with the mechanism that lead to ${\cal
M}_{\nu,33}$ becoming small as a result of $b-\tau$ convergence. Hence the
natural way that $\theta_A$ became large in the CP conserving case
remains.

Let us discuss if the new model is still predictive in the neutrino
sector. Of the three new parameters, one is determined by the CP violating
quark phase. the two others are determined by the solar mixing angle and
the solar mass difference squared. Therefore we lose the prediction for
these parameters. However, we can predict in addition to $\theta_A$ which 
is close to maximal, $\theta_{13}\geq 0.1$ and the Dirac phase for the
neutrinos. We show the predictions for Dirac phase in Fig. 4. This is a
unique property of the model that it can predict the leptonic CP phase.

\begin{figure}[p]
\centering
\includegraphics*[angle=0,width=11cm]{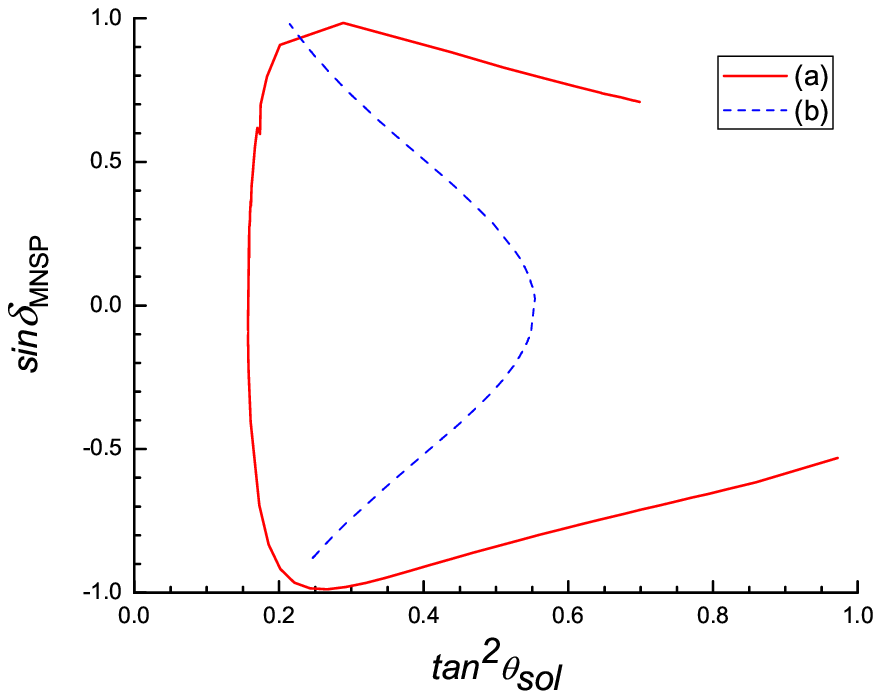}
 \caption{The prediction of MNSP phase is plotted as a function of the
solar mixing angle. The  two lines (a) and (b) correspond to two choices
of signs of the fermion masses. The phases are plotted for
$\Delta m^2_{sol}/\Delta m^2_A = 0.02$.}
\label{Fig.4}
\end{figure}

\section{Radiative generation of large mixings: another application of
type II seesaw}
As alluded before, type II seesaw liberates the neutrinos from obeying
normal generational hierarchy and instead could easily be quasi-degenerate
in mass. This raises a new way to understand the large mixings instead of
having to generate them in the original seesaw theory as is normally done.
 The basic idea is that at the seesaw
scale, all mixings angles are small.
 Since the observed neutrino mixings are the weak scale
observables, one must extrapolate\cite{babu1} the seesaw scale mass
matrices to the weak scale and recalculate the mixing angles.
The extrapolation formula is
\begin{eqnarray}
{ M}_{\nu}(M_Z)~=~ {\bf I}{ M_{\nu}} (v_R) {\bf I}\\
where~~~~~~~~{\bf I}_{\alpha \alpha}~=~
\left(1-\frac{h^2_{\alpha}}{16\pi^2}\right)
\end{eqnarray}
Note that since $h_{\alpha}= \sqrt{2}m_{\alpha}/v_{wk}$ ($\alpha$ being
the charged lepton index), in the extrapolation only the $\tau$-lepton
makes a difference. In the MSSM, this increases the ${ M}_{\tau\tau}$
entry of the neutrino mass matrix and essentially leaves the others
unchanged. It was shown
in ref.\cite{balaji} that if the muon and the tau neutrinos are
nearly degenerate but not degenerate enough in mass at the seesaw scale,
the radiative corrections can become large enough so that at the weak
scale the two diagonal elements of ${ M}_{\nu}$
 become much more degenerate. This leads to an enhancement of the
mixing angle to become almost maximal value.
This can also be seen from the renormalization group equations when they
are written in the mass basis\cite{casas}. Denoting the mixing angles as
$\theta_{ij}$ where $i,j$ stand for generations, the equations are:
\noindent
\begin{eqnarray}
\dsodt&=&-F_{\tau}{c_{23}}^2\left(
-s_{12}U_{\tau1}D_{31}+c_{12}U_{\tau2}D_{32}
\right),\label{eq3}\\
\dstdt&=&-F_{\tau}c_{23}{c_{13}}^2\left(
c_{12}U_{\tau1}D_{31}+s_{12}U_{\tau2}D_{32}
\right),\label{eq4}\\
\dsthdt&=&-F_{\tau}c_{12}\left(c_{23}s_{13}s_{12}U_{\tau1}
D_{31}-c_{23}s_{13}c_{12}U_{\tau2}D_{32}\right.\nonumber \\
&&\left.+U_{\tau1}U_{\tau2}D_{21}\right).\label{eq5}\end{eqnarray}
\noindent
where $D_{ij}={\left(m_i+m_j)\right)/\left(m_i-m_j\right)}$ and
$U_{\tau 1,2,3}$ are functions of the neutrino mixings angles. The
presence
of $(m_i-m_j)$ in the denominator makes it clear that as $m_i\simeq m_j$,
that particular coefficient becomes large and as we extrapolate from the
GUT scale to the weak scale, small mixing angles at GUT scale become large
at the weak scale.
It has been shown recently that indeed such a mechanism for understanding
large mixings can
work for three generations\cite{par}. It was shown that if we identify
the seesaw scale
neutrino mixing angles with the corresponding quark mixings and assume
quasi-degenerate neutrinos, the weak scale solar
and atmospheric angles get magnified to the desired level while due to the
extreme smallness of $V_{ub}$, the magnified value of $U_{e3}$ remains
within its present upper limit. In figure 5, we show the evolution of the
mixing angles to the weak scale.
 \begin{figure}
\epsfxsize=8.5cm
\epsfbox{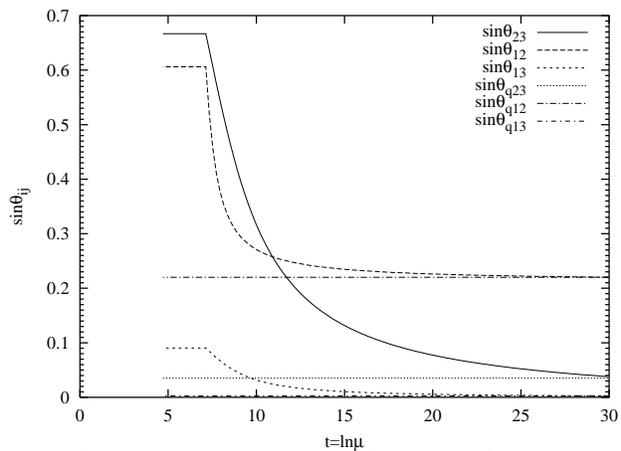}
\caption{Radiative magnification of small quark-like neutrino mixings at
the see-saw scale to bilarge values at low energies. The solid, dashed and
dotted lines represent
$\sin\theta_{23}$, $\sin\theta_{13}$, and $\sin\theta_{12}$,
respectively.}
\label{fig1}
\end{figure}
A requirement for this scenario to work is that the common mass of
neutrinos must be larger than $0.1$ eV, a result that can be tested in
neutrinoless double beta experiments.

\section{Quark-lepton complementarity and large solar mixing}
There has been a recent suggestion\cite{raidal} that perhaps the large
but not maximal solar mixing angle is related to physics of the quark
sector. According to this, the deviation
from maximality
of the solar mixing may be related to the quark mixing angle
$\theta_C\equiv \theta^{q}_{12}$ and is based on the
observation that the mixing angle responsible
for solar neutrino oscillations, $\theta_{\odot}\equiv \theta^\nu_{12}$
satisfies an interesting
complementarity relation with the corresponding angle in the quark sector
$\theta_{Cabibbo}\equiv \theta^q_{12}$ i.e. $\theta^\nu_{12}+\theta^q_{12}
\simeq \pi/4$.
While it
is quite possible that this relation is purely accidental or due to some
other dynamical effects, it is interesting
to pursue the possibility that there is a deep meaning behind it
and see where it leads. It has been shown in a recent paper that if
Nature is quark lepton unified at high scale, then a relation between
$\theta^\nu_{12}$ and $\theta^q_{12}$ can be obtained in a natural manner
provided the neutrinos obey the inverse hierarchy\cite{fram}. It predicts
$sin^2\theta_\odot\simeq 0.34$ which agrees with present data at the
2$\sigma$ level. It also predicts a large $\theta_{13}\sim 0.18$, both of
which are predictions that can be tested experimentally in the near
future.

\section{Conclusion}
In summary, the seesaw mechanism is by far the simplest and most
appealing way to understand neutrino masses. It not only improves
the aesthetic appeal of the standard model by restoring
quark-lepton symmetry but it also makes weak interactions
asymptotically parity conserving. Further more it connects
neutrino masses with the hypothesis of grand unification. In this
talk I have discussed three ways to understand the large solar
and atmospheric neutrino mixings within the frameworks that unify
quarks and lepton and in one case into a grand unified model
based on SO(10). All three models predict large values for
$\theta_{13}$ and can therefore be tested in forthcoming
experiments. The SO(10) model appears to be  most promising since it
not only resolves the difficulties of the minimal SUSY SU(5) GUT but
is also a minimal predictive model for neutrinos.

 From these examples, one is also tempted to conclude
that a large $\theta_{13}$ could be a generic feature of models
that unify quarks and leptons, which if true will be a unique
window to a very important question in beyond the standard model
physics.

 This work is partially supported by the National Science Foundation
Grant No. PHY-0354401. I would like to thank the organizers of the
Nobel symposium 129 at Haga Slott for creating a very pleasant
environment for physics.


\begin{thebibliography}{99}

\bibitem{seesaw}  P. Minkowski, Phys. lett. {\bf B67 }, 421
(1977); M.~Gell-Mann, P.~Ramond, and R.~Slansky, \emph{Supergravity}
(P.~van Nieuwenhuizen et al. eds.), North Holland, Amsterdam, 1980, p.~315;
 T.~Yanagida, in \emph{Proceedings of the
Workshop on the
Unified Theory and the Baryon Number in the Universe} (O.~Sawada
and A.~Sugamoto, eds.), KEK, Tsukuba, Japan, 1979, p.~95; S.~L.
Glashow, \emph{The future of elementary particle physics}, in
  \emph{Proceedings of the 1979 Carg{\`e}se Summer Institute on Quarks and
  Leptons} (M.~L{\'e}vy et al. eds.), Plenum Press, New York, 1980,
pp.~687--713;  R.~N. Mohapatra and G.~Senjanovi{\'c},
Phys. Rev. Lett. \textbf{44} 912 (1980).


\bibitem{rev} For recent reviews, see J.~N.~Bahcall, M.~C.~Gonzalez-Garcia
and C.~Pena-Garay, JHEP {\bf 0408}, 016 (2004)
[arXiv:hep-ph/0406294]; C. Gonzalez-Garcia and M. Maltoni,
hep-ph/0406056; M. Maltoni, T. Schwetz, M. Tortola and J. W. F. Valle,
hep-ph/0405172; A. de Gouvea, hep-ph/0411274.


\bibitem{reactor} K.~Anderson {\it et al.},
arXiv:hep-ex/0402041; M.~Apollonio {\it et al.}, Eur.\ Phys.\ J.\
C {\bf 27}, 331 (2003) arXiv:hep-ex/0301017.

\bibitem{lbl} M.~V.~Diwan {\it et al.},
Phys.\ Rev.\ D {\bf 68}, 012002
(2003) arXiv:hep-ph/0303081; D. Ayrea et al. hep-ex/0210005;
Y. Itow et al. (T2K collaboration) hep-ex/0106019; I.~Ambats {\it et al.}
(NOvA Collaboration), FERMILAB-PROPOSAL-0929.

\bibitem{so10}  H. Georgi, in {\it Particles and Fields}
(ed. C. E. Carlson), A. I. P. (1975);
H. Fritzsch and P. Minkowski, Ann. of Physics, Ann. Phys. {\bf 93}, 193
(1975).

\bibitem{ps} J. C. Pati and A. salam, Phys. Rev. {\bf D10}, 275 (1974).

\bibitem{janet} J. Conrad, these proceedings.

\bibitem{lrs} J. C. Pati and A. Salam, \cite{ps};
R. N. Mohapatra and J. C. Pati, Phys. Rev. {\bf D 11}, 566, 2558 (1975);
G. Senjanovi\'c and R. N. Mohapatra, Phys. Rev. {\bf D 12}, 1502 (1975).

\bibitem{marshak} R. N. Mohapatra and R. E. Marshak, Phys. Lett. {\bf B
91}, 222 (1980);
A. Davidson, Phys. Rev. {\bf D20}, 776 (1979).

\bibitem{seesaw2}  G. Lazarides, Q. Shafi and C. Wetterich,
Nucl.Phys.{\bf B181}, 287 (1981); R. N. Mohapatra and G. Senjanovi\'c,
Phys. Rev. {\bf D 23}, 165 (1981);

\bibitem{king} A. Smirnov, hep-ph/0311259; S. F. King,
Rept.Prog.Phys. {\bf 67}, 107  (2004); G. Altarelli and F. Feruglio,
hep-ph/0405048; R. N. Mohapatra, hep-ph/0211252; New J. Phys., {\bf 6}, 82
(2004).

\bibitem{caldwell}  See for instance D.~O.~Caldwell and R.~N.~Mohapatra,
Phys.\ Rev.\ D {\bf 48}, 3259 (1993); D.~G.~Lee and R.~N.~Mohapatra,
Phys.\ Lett.\ B {\bf 329}, 463 (1994); S.~Antusch and S.~F.~King,
hep-ph/0402121.

%
\bibitem{mutau} R. N. Mohapatra, Slac Summer
Inst. lecture; http://www-conf.slac.stanford.edu/ssi/2004; hep-ph/0408187;
JHEP, {\bf 10}, 027 (2004);
  W. Grimus, A. S.Joshipura, S. Kaneko, L.
Lavoura, H. Sawanaka, M. Tanimoto, hep-ph/0408123.

\bibitem{babu} K. S. Babu and R. N. Mohapatra,
Phys. Rev. Lett. {\bf 70}, 2845 (1993).

\bibitem{last} D. G. Lee and R. N. Mohapatra, Phys. Rev. {\bf D 51}, 1353
(1995); L. Lavoura, Phys. Rev. {\bf D 48}, 5440 (1993); B. Brahmachari and
R. N. Mohapatra, Phys. Rev. {\bf D58}, 015001 (1998); K. Oda, E. Takasugi,
M. Tanaka and M. Yoshimura, Phys. Rev. {\bf D 58}, 055001 (1999);
K. Matsuda, Y. Koide, T. Fukuyama and N. Okada, Phys. Rev. {\bf D
65}, 033008 (2002); T. Fukuyama and N. Okada, hep-ph/0205066; N. Oshimo,
hep-ph/0305166.

\bibitem{goran}  B.~Bajc, G.~Senjanovic and F.~Vissani,
Phys.\ Rev.\ Lett.\  {\bf 90}, 051802 (2003)
[arXiv:hep-ph/0210207]

\bibitem{goh} H.~S.~Goh, R.~N.~Mohapatra and S.~P.~Ng,
Phys.\ Lett.\ B {\bf 570}, 215 (2003)
[arXiv:hep-ph/0303055].

\bibitem{more} C.~S.~Aulakh, B.~Bajc, A.~Melfo, G.~Senjanovic and
F.~Vissani,
Phys.\ Lett.\ B {\bf 588}, 196 (2004)
[hep-ph/0306242]; C. S. Aulakh and A. Giridhar, hep-ph/0204097;
 T.~Fukuyama, A.~Ilakovac, T.~Kikuchi, S.~Meljanac and
N.~Okada, arXiv:hep-ph/0401213;
S.~Bertolini, M.~Frigerio and M.~Malinsky,
hep-ph/0406117;
B.~Dutta, Y.~Mimura and R.~N.~Mohapatra, hep-ph/0406262;
W.~M.~Yang and Z.~G.~Wang, hep-ph/0406221;
H.~S.~Goh, R.~N.~Mohapatra, S.~Nasri and S.~P.~Ng,
Phys.\ Lett.\ B {\bf 587}, 105 (2004); H.~S.~Goh, R.~N.~Mohapatra and
S.~Nasri, arXiv:hep-ph/0408139; Phys. Rev. {\bf D 70}, 075002 (2004).

\bibitem{mimura} B. Dutta, Y. Mimura and R. N. Mohapatra, hep-ph/0406262;
Phys. Lett. {\bf B} (2004).

\bibitem{babu1}  K.S.~Babu, C.N.~Leung and J.~Pantaleone, Phys.~Lett.~{\bf
B319}, 191 (1993); P. Chankowski and Z. Pluciennik, Phys. Lett. {\bf
B316}, 312 (1993).

\bibitem{balaji} K. S. Balaji, A. Dighe, R. N. Mohapatra and M. K. Parida,
Phys. Rev. Lett. {\bf 84}, 5034 (2000); Phys. Lett. {\bf B481 }, 33
(2000).

\bibitem{casas}  J. Casas, J. Espinoza, J. Ibara and S. Navaro,
hep-ph/9905381; Nucl. Phys. {\bf B 573}, 652 (2000).

\bibitem{par} R. N. Mohapatra, M. K. Parida and G. Rajasekaran,
hep-ph/0301234;  Phys.Rev.{\bf D69}, 053007 (2004).


\bibitem{raidal} M. Raidal, hep-ph/0404046; H. Minakata and A. Y. Smirnov.
{\tt hep-ph/0405088}.

\bibitem{fram} P. Frampton and R. N. Mohapatra, hep-ph/0407139.

\end{thebibliography}
\end{document}